\documentclass[10pt,twocolumn,conference]{IEEEtran}

\IEEEoverridecommandlockouts

\usepackage{graphicx}
\usepackage{amsmath}
\usepackage{amssymb}
\usepackage[caption=false]{subfig}
\usepackage[noadjust]{cite}
\usepackage{float}
\usepackage{algorithm}
\usepackage{bibentry}
\usepackage{balance}
\usepackage{algorithm}
\usepackage{algorithmicx}
\usepackage{algpseudocode}
\usepackage{xcolor}
\usepackage{subfig}
\graphicspath{{img/}}

\begin{document}

\bstctlcite{IEEEexample:BSTcontrol}

\title{Power Allocation for Fingerprint-Based PHY-Layer Authentication with mmWave UAV Networks\thanks{This work is supported in part by the INL Laboratory Directed Research Development (LDRD) Program under DOE Idaho Operations Office Contract DEAC07-05ID14517.}}

\author{
\IEEEauthorblockN{Sung Joon Maeng$^*$, Yavuz Yap{\i}c{\i}$^\dagger$, \.{I}smail G\"{u}ven\c{c}$^*$, Huaiyu Dai$^*$, and Arupjyoti Bhuyan$^{\dagger\dagger}$}\IEEEauthorblockA{$^*$Department of Electrical and Computer Engineering, North Carolina State University, Raleigh, NC\\
$^\dagger$Electrical Engineering Department, University of South Carolina, Columbia, SC\\
$^{\dagger\dagger}$Idaho National Laboratory, Idaho Falls, ID\\
\{smaeng, iguvenc, hdai\}@ncsu.edu, 	yyapici@mailbox.sc.edu, arupjyoti.bhuyan@inl.gov}}

\maketitle

\begin{abstract}
Physical layer security (PLS) techniques can help to protect wireless networks from eavesdropper attacks.  In this paper, we consider the authentication technique that uses fingerprint embedding to defend 5G cellular networks with unmanned aerial vehicle (UAV) systems from eavesdroppers and intruders. Since the millimeter wave (mmWave) cellular networks use narrow and directional beams, PLS can take further advantage of the 3D spatial dimension for improving the authentication of UAV users. Considering a multi-user mmWave cellular network, we propose a power allocation technique that jointly takes into account splitting of the transmit power between the precoder and the authentication tag, which manages both the secrecy as well as the achievable rate. Our results show that we can obtain optimal achievable rate with expected secrecy.
\end{abstract}

\begin{IEEEkeywords}
    5G, authentication, artificial noise, eavesdropping, fingerprinting, millimeter-wave, physical layer security, power allocation, precoder.
\end{IEEEkeywords}

\section{Introduction}
As the demand for wireless data continues to increase, millimeter-Wave (mmWave) communication is emerging as a core feature of future mobile networks, taking advantage of very large bandwidths for dramatically increasing the data rates~\cite{xiao2017millimeter}. In parallel, use of unmanned aerial vehicles (UAV) have been finding various use cases, from delivery to inspection, entertainment, and public safety~\cite{hayat2016survey}. To support such use cases, maintaining reliable and secure beyond visual line-of-sight (BVLOS) connectivity with UAVs is becoming a critical challenge~\cite{8891595}. Cellular networks, with their widely deployed base station (BS) infrastructure, can help sustain such reliable and secure BVLOS links with UAVs. 

Since UAV-mounted networks constitute heterogeneous architecture and are vulnerable to be wiretapped by ground unauthorized parties, ensuring secure communications with UAVs is critical~\cite{li2019secure,maeng2020precoder}. Traditionally, cryptographic encryption is implemented at the network and the application layer for achieving secure transmissions. During the past decade, physical layer security (PLS) has received significant attention as a promising enhancement to traditional security techniques~\cite{yang2015safeguarding}.
In a  wire-tap channel, 
the transmitter desires to transmit data securely to the legitimate user and aims to minimize the information leakage to an eavesdropper as much as possible~\cite{wyner1975wire}. To achieve that, artificial noise (AN) can be superposed on the data transmission in  multiple-input-multiple-output (MIMO) system~\cite{goel2008guaranteeing,zhu2014secure}. For example, in the precoder design of~\cite{zhu2014secure}, power splitting factor is introduced to divide the allocated power to data and AN precoders, and the optimal power allocation that maximizes the ergodic secrecy rate is studied.

Authentication is another pivotal aspect of PLS. Commonly, hash-based message authentication codes (HMAC) are used for the authentication procedure. However, HMAC is vulnerable when eavesdropper has noise-less access. Fingerprint embedded authentication is a promising technique which has been proposed in~\cite{paul2015wireless,paul2011mimo}. In this technique, slightly power allocated authentication tag is superimposed on data, which decreases the ability of the adversary to decode authentication tag as well as does not require additional bandwidth. The AN in the fingerprinting authentication framework is introduced on multiple-input-single-output (MISO) channel in \cite{perazzone2019fingerprint}. The joint data and tag precoder is optimized in \cite{chalise2019joint}.

\begin{figure}[!t]
	\centering
	\vspace{-0.0in}
	\includegraphics[width=0.38\textwidth]{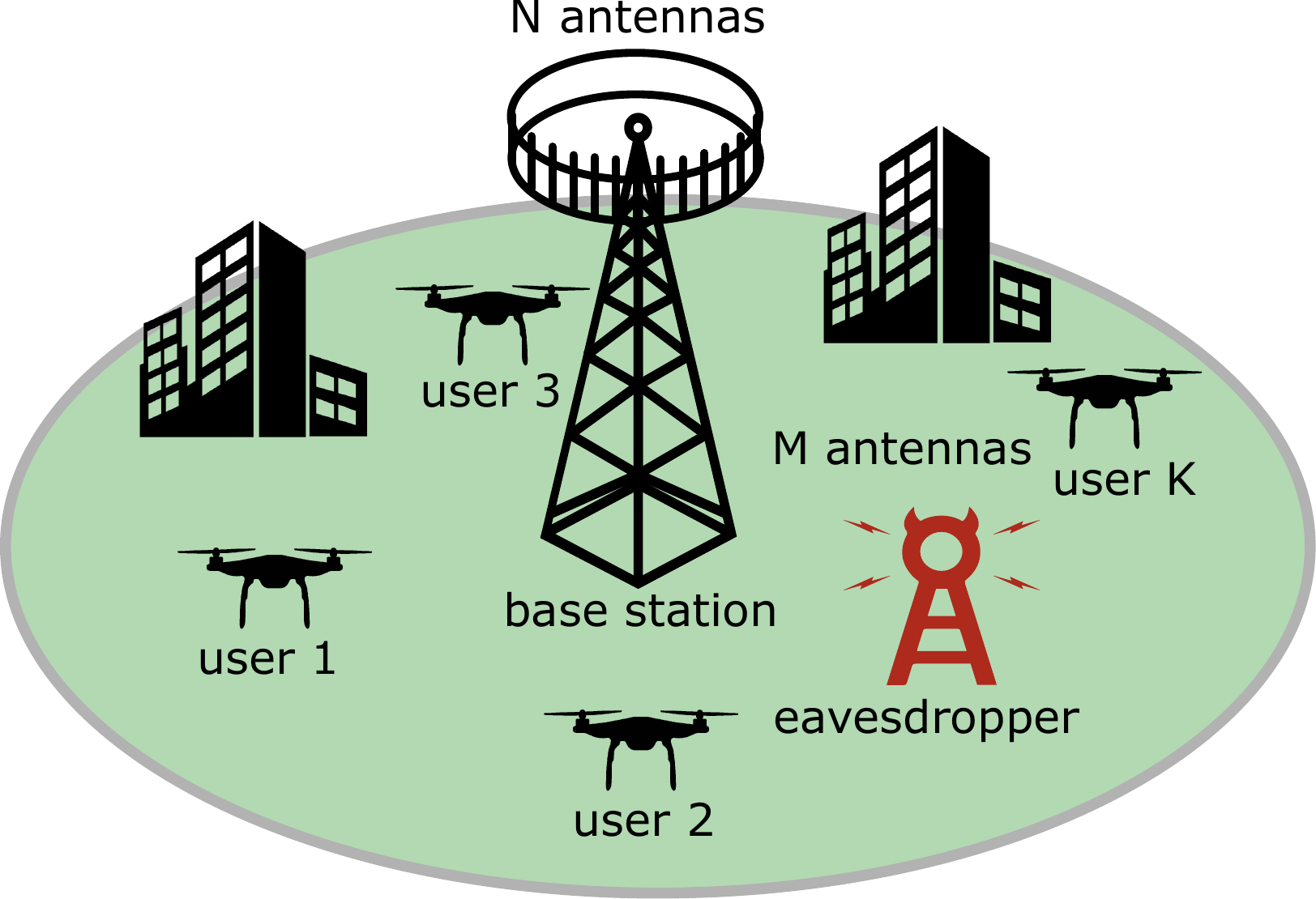}
	\vspace{-0in}
	\caption{System model with cellular-connected UAVs and mmWave links.}
	\label{fig:illu}
	\vspace{-0.1in}
\end{figure}

In this paper, we consider a wire-tap channel multi-user scenario with cellular-connected mmWave UAVs, as illustrated in Fig.~\ref{fig:illu}.  We also consider AN transmission and fingerprint embedded authentication procedure. In this setup, we propose power allocation strategies, which manages the authentication success probability of the legitimate users as well as the authentication key deciphering  probability of the eavesdropper. Unlike the related works discussed earlier, which optimize either the power between data and AN precoder or the power of authentication tag without AN, we jointly take into account 1) the allocated power of the data and the AN,  and 2)  authentication tag power. Using simulation results, we show that the proposed strategy can strike a desired balance between the sum rate,  authentication probability, and the expected decoder success probability. 

\begin{figure}[!t]
	\centering
	\vspace{-0.0in}
	\includegraphics[width=0.35\textwidth]{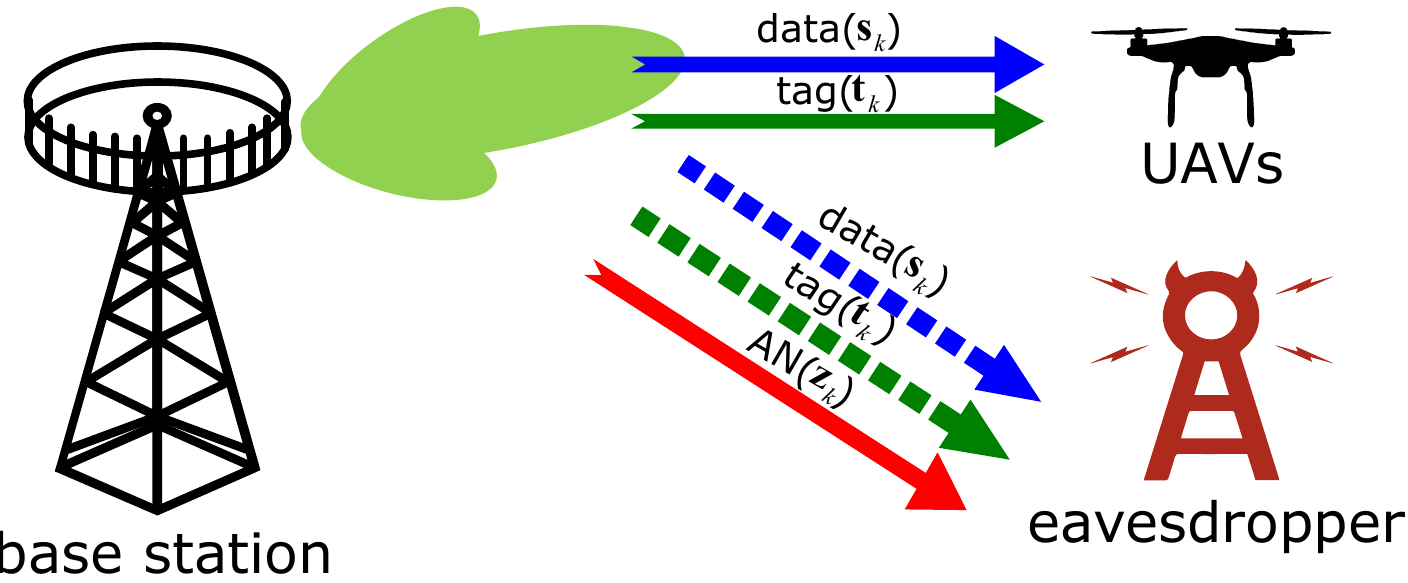}
	\vspace{-0.0in}
	\caption{System model with data, authentication tag, and artificial noise transmission. Power of the data transmission (including the tag) is $\phi \mathsf{P}_{\rm Tx}$, while the power of the AN transmission is $(1-\phi)\mathsf{P}_{\rm Tx}$}.
	\label{fig:illu2}
	\vspace{-0.2in}
\end{figure}

\section{System Model} \label{sec:system}

We consider a mmWave communications scenario where a single BS serves $K$ UAVs (users), each with single antenna, as shown in Fig.~\ref{fig:illu}. The cellular network is wiretapped by a single passive eavesdropper (Eve), which illegally listens the transmitted messages. The BS and the Eve are equipped with $N$ and $M$ elements uniform linear array (ULA) antenna. We consider multi-user MIMO (MU-MIMO) system where all users share the same time and frequency resources. For the sake of increasing the resiliency against potential attacks, the BS broadcasts artificial noise (AN) superposed on the data. In addition, for authentication purpose, the BS superimposes the encrypted tag ($\textbf{t}_k$) for the separate users' data symbols ($\textbf{s}_k$). Fig.~\ref{fig:illu2} shows the system model of data, tag, and artificial noise transmission.

Then, the received signal at the $u$-th user is given by 
\begin{align}\label{eq:user_y}
    \textbf{y}_u^{\rm H}&=\sum_{k=1}^{K}\sqrt{\mathsf{P}_\mathsf{Tx}}\textbf{h}_u^{\rm H}\textbf{w}_k(\sqrt{\mathsf{P}_\mathsf{s}}\textbf{s}_k+\sqrt{\mathsf{P}_\mathsf{t}}\textbf{t}_k)^{\rm H}\nonumber\\
    &+\sum_{i=1}^{Z}\sqrt{\mathsf{P}_\mathsf{Tx}}\textbf{h}_u^{\rm H}\textbf{v}_i\textbf{z}_i^{\rm H}+\textbf{n}_u^{\rm H},
\end{align}
where $\mathsf{P}_\mathsf{Tx}$ is the transmit power of the BS, $\textbf{h}_u\in\mathbb{C}^{N\times 1}$ is the $u$-th user's channel vector, $\textbf{W}=[\textbf{w}_1,\dots,\textbf{w}_K]\in\mathbb{C}^{N\times K}$ is the data precoder matrix whose $\textbf{w}_k$ is the precoder vector for the $k$-th user, $\mathsf{P}_\mathsf{s}$ and $\mathsf{P}_\mathsf{t}$ are allocated power to data symbols and tag with power being normalized by 1 ($\mathsf{P}_\mathsf{s} + \mathsf{P}_\mathsf{t}=1$),  $\textbf{V}=[\textbf{v}_1,\dots,\textbf{v}_K]\in\mathbb{C}^{N\times Z}$ is the AN precoder matrix,  and $\textbf{n}_u$ is additive noise with $\mathcal{CN}(0, \sigma_n^2)$. We assume that data symbols ($\textbf{s}_k\in\mathbb{C}^{L_{\mathsf{t}}\times 1}$), tag ($\textbf{t}_k\in\mathbb{C}^{L_{\mathsf{t}}\times 1}$), and AN ($\textbf{z}_i\in\mathbb{C}^{L_{\mathsf{t}}\times 1}$) follow $\mathcal{CN}(0,1)$. 

The channel vector of the individual users are modeled by the mmWave channel model, the path of which is represented by the angle of departure ($\theta_{k,l}$), as
\begin{align}\label{eq:channel_user}
    \textbf{h}_k = \sqrt{ \frac{ N }{ L_\mathsf{p} } } \sum^{L_\mathsf{p}}_{\ell=1} \frac{\alpha_{k,\ell}}{\mathsf{PL} \left( d_k \right)}  \textbf{a}_N(\theta_{k,\ell}),
\end{align}
where $L_\mathsf{p}$ is the number of paths, $\alpha_{k,\ell}$ is the small scale fading factor which is modeled by the unit complex Gaussian. $\mathsf{PL} \left( d_k \right)$ is the pathloss by distance $d_k=\sqrt{d_{\mathsf{V},k}^2+d_{\mathsf{H},k}^2}$, which is given by the 3GPP urban micro (UMi) as:
\begin{align}\label{eq:PL}
    \mathsf{PL} \left( d_k \right) &= 32.4 + 21\log_{10}\left( d_k \right) + 20\log_{10}\left( f_{\rm c} \right),
\end{align}
where $f_{\rm c}$ is carrier frequency. Moreover,
$\textbf{a}_N(\theta_{k,\ell})$ is the array steering vector of $\ell$-th path of $k$-th user, which is given by
\begin{align}\label{eq:steering_vector}
    \textbf{a}_N(\theta)&=\frac{1}{\sqrt{N}}\left[1\; e^{-j\frac{2\pi d_{\rm s}}{\lambda}\sin\theta}\;\dots\; e^{-j\frac{2\pi d_{\rm s}}{\lambda}(N-1)\sin\theta}\right]^{\rm T},
\end{align}
where $d_{\rm s}$, $\lambda$ are the antenna spacing between the elements and the wavelength. Note that we assume that the angle of departure ($\theta$) follows Laplacian distribution around the line-of-sight (LoS) angle $\theta_\mathsf{LoS}$ with the angle spread of $\Delta$. 

On the other hand, Eve's received signal is expressed as
\begin{align}
    \textbf{Y}_{\rm e}^{\rm H}&=\sum_{k=1}^{K}\sqrt{\mathsf{P}_\mathsf{Tx}}\textbf{H}_{\rm e}^{\rm H}\textbf{w}_k(\sqrt{\mathsf{P}_\mathsf{s}}\textbf{s}_k+\sqrt{\mathsf{P}_\mathsf{t}}\textbf{t}_k)^{\rm H}\nonumber\\
    &+\sum_{i=1}^{Z}\sqrt{\mathsf{P}_\mathsf{Tx}}\textbf{H}_{\rm e}^{\rm H}\textbf{v}_i\textbf{z}_i^{\rm H}+\textbf{N}_{\rm e}^{\rm H}~,
\end{align}
where $\textbf{H}_\mathsf{e}\in\mathbb{C}^{N\times M}$ is the channel matrix of Eve, and $\textbf{N}_\mathsf{e}\in\mathbb{C}^{M\times L_{t}}$ is the additive noise following $\mathcal{CN}(0, \sigma_n^2)$. The channel matrix between the BS and the Eve ($\textbf{H}_{\rm e}$) can be written as
\begin{align}\label{eq:channel_eve}
    \textbf{H}_\mathsf{e} = \sqrt{\frac{NM}{L_\mathsf{p}}} \sum^{L_\mathsf{p}}_{\ell=1} \frac{ \alpha_{\mathsf{e},\ell} }{ \mathsf{PL} \left( d_\mathsf{e} \right) } \textbf{a}_N(\theta_{\mathsf{e},\ell})\textbf{a}_M^{\rm H}(\zeta_{\mathsf{e},\ell}),
\end{align}
where $\zeta_{\mathsf{e},\ell}$ is the angle of arrival of the $\ell$-th path associated with the Eve, the distribution of which follows the same distribution as the angle of departure.

We evaluate the performance of the system by the sum rate of the legitimate users. The SINR is given by
\begin{align}\label{eq:sum_rate}
     \!\!\mathsf{SINR}_u = \frac{\mathsf{P}_\mathsf{s}\textbf{w}_u^{\rm H}\textbf{h}_u\textbf{h}_u^{\rm H}\textbf{w}_u}{\sum_{k\neq u}^{K}\mathsf{P}_\mathsf{s} \textbf{w}_k^{\rm H}\textbf{h}_u\textbf{h}_u^{\rm H}\textbf{w}_k + \sum_{i=1}^{Z} \textbf{v}_i^{\rm H}\textbf{h}_u\textbf{h}_u^{\rm H}\textbf{v}_i + \rho^{{-}1}},
\end{align}
where $\rho \,{=}\, \frac{ \mathsf{P}_\mathsf{Tx}}{ \sigma_n^2}$ is the signal-to-noise ratio (SNR). Then, the sum rate of the legitmated users can be written as
\begin{align}\label{eq:sum_rate}
    R_{\rm sum}&=\sum_{k=1}^K\log_2\left(1+
    \mathsf{SINR}_k\right).
\end{align}

\section{Fingerprint Authentication for Users and Eve}

In this section, we describe the authentication and the impersonation procedure for the legitimate user and Eve. The BS shares the predetermined key ($\eta_u$) with the legitimate user for the authentication on the user side. The BS generates the tag from the key and data symbols associated with the legitimate user. The tag is generated by
\begin{align}\label{tag_gen}
    \textbf{t}_u&=f(\textbf{s}_u,\eta_u)~,
\end{align}
where the tag generation function $f(\cdot)$ is a one-way, collision-resistant function \cite{paul2011mimo}. The BS allocates the small portion of power to the generated tag and transmits the tag overlapped on data symbols. The transmitted power of tag occupies only small percentage of the power of the data symbols so that the SNR from the legitimate user is merely affected. In following subsections, we elaborate the authentication procedure by detecting the tag in both the legitimate user and the Eve sides, and express the authentication success probability and the correct key detecting probability, respectively.

\subsection{Authentication Procedure for Legitimate User}

In this paper, the authentication is proceeded with the assumption of the correct detection of the data symbols ($\textbf{s}_u$). Note that the tag generation in \eqref{tag_gen} requires date symbols, and the legitimate user successfully regenerates the tag only with the correct recovered data symbols. For the simple notation, we indicate the legitimate user as $u$-th user. Then, the received signal of the legitimate user in \eqref{eq:user_y} can be expressed after subtracting the contribution of the decoded data as 
\begin{align}
    \textbf{r}_u^{\rm H}&=\sqrt{\mathsf{P}_\mathsf{Tx}\mathsf{P}_\mathsf{t}}\textbf{h}_u^{\rm H}\textbf{w}_u\textbf{t}_u^{\rm H}+\sum_{k\neq u}\sqrt{\mathsf{P}_\mathsf{Tx}}\textbf{h}_u^{\rm H}\textbf{w}_k(\sqrt{\mathsf{P}_\mathsf{s}}\textbf{s}_k+\sqrt{\mathsf{P}_\mathsf{t}}\textbf{t}_k)^{\rm H}\nonumber\\
    &+\sum_{i=1}^{Z}\sqrt{\mathsf{P}_\mathsf{Tx}}\textbf{h}_u^{\rm H}\textbf{v}_i\textbf{z}_i^{\rm H}+\textbf{n}_u^{\rm H}.
\end{align}
After equalizing the contribution of the tag by the least square (LS) method, we can represent the residual signal as the estimated tag as follows: 
\begin{align}
    \hat{\textbf{t}}_u^{\rm H}&=\textbf{t}_u^{\rm H}+\underbrace{\sum_{k\neq u}\frac{\textbf{h}_u^{\rm H}\textbf{w}_k}{\sqrt{\mathsf{P}_\mathsf{t}}\textbf{h}_u^{\rm H}\textbf{w}_u}(\sqrt{\mathsf{P}_\mathsf{s}}\textbf{s}_k+\sqrt{\mathsf{P}_\mathsf{t}}\textbf{t}_k)^{\rm H}}_\text{multiuser interference}\nonumber\\
    &+\underbrace{\sum_{i=1}^{Z}\frac{\textbf{h}_u^{\rm H}\textbf{v}_i}{\sqrt{\mathsf{P}_\mathsf{t}}\textbf{h}_u^{\rm H}\textbf{w}_u}\textbf{z}_i^{\rm H}}_\text{artificial noise}+\underbrace{\frac{1}{\sqrt{\mathsf{P}_\mathsf{Tx}\mathsf{P}_\mathsf{t}}\textbf{h}_u^{\rm H}\textbf{w}_u}\textbf{n}_u^{\rm H}}_\text{additive noise},
\end{align}
where the estimated tag is corrupted by multiuser interference, AN, and additive noise from the receiver circuits. With the shared key from the BS, the legitimate user generates the expected key along with the recovered data symbols:
\begin{align}
    \Tilde{\textbf{t}}_u&=f(\hat{\textbf{s}}_u,\eta_u).
\end{align}
Then, the legitimate user carries out hypothesis test, which authenticates the user. The legitimate user calculates the correlation between the expected tag and the estimated tag, as
\begin{align}
    \tau_b&=\Re(\hat{\textbf{t}}_u^{\rm H}\Tilde{\textbf{t}}_u).
\end{align}
By comparing the correlation with the threshold value ($\tau_0$), we can determine the hypotheses,
\begin{align*}
    H_0:& \text{ not authentic,}\quad\tau_b\leq\tau_0\\
    H_1:& \text{ authentic,} \quad\tau_b>\tau_0.
\end{align*}

With the assumption of the sufficiently large length of tag symbols ($L_{\mathsf{t}}$), the distribution of $\tau_b$ can be approximated as Gaussian by the central limit theorem. Then, we can derive the mean and the variance of $\tau_b$ on both hypotheses as:
\begin{align}
    \mathbb{E}\{\tau_b|H_0\}&=\mu_{u,0}=0,\label{eq:H_0_m_1}\\
    \text{Var}\{\tau_b|H_0\}&=\sigma_{u,0}^2=\frac{L_{\rm t}}{2}\left(1+\sum_{k\neq u}\frac{|\textbf{h}_u^{\rm H}\textbf{w}_k|^2}{\mathsf{P}_\mathsf{t}|\textbf{h}_u^{\rm H}\textbf{w}_u|^2}\right.\nonumber\\
    &\left.+\sum_{i=1}^{Z}\frac{|\textbf{h}_u^{\rm H}\textbf{v}_i|^2}{\mathsf{P}_\mathsf{t}|\textbf{h}_u^{\rm H}\textbf{w}_u|^2}+\frac{1}{\mathsf{P}_\mathsf{Tx}\mathsf{P}_\mathsf{t}|\textbf{h}_u^{\rm H}\textbf{w}_u|^2}\sigma_n^2\right),\label{eq:H_0_v_1}\\
    \mathbb{E}\{\tau_b|H_1\}&=\mu_{u,1}=L_{\rm t},\label{eq:H_1_m_1}\\
    \text{Var}\{\tau_b|H_1\}&=\sigma_{u,1}^2=\frac{L_{\rm t}}{2}\left(2+\sum_{k\neq u}\frac{|\textbf{h}_u^{\rm H}\textbf{w}_k|^2}{\mathsf{P}_\mathsf{t}|\textbf{h}_u^{\rm H}\textbf{w}_u|^2}\right.\nonumber\\
    &\left.+\sum_{i=1}^{Z}\frac{|\textbf{h}_u^{\rm H}\textbf{v}_i|^2}{\mathsf{P}_\mathsf{t}|\textbf{h}_u^{\rm H}\textbf{w}_u|^2}+\frac{1}{\mathsf{P}_\mathsf{Tx}\mathsf{P}_\mathsf{t}|\textbf{h}_u^{\rm H}\textbf{w}_u|^2}\sigma_n^2\right)\label{eq:H_1_v_1}.
\end{align}
By utilizing above derived statistics, we can obtain the authentication probability for a given false alarm probability ($p_{\rm fa}$). The threshold value can be written as
\begin{align}
    \tau_0&=\Phi^{-1}(1-p_{\rm fa})\sigma_{u,0},
\end{align}
where $\Phi(\cdot)$ is the CDF of the Gaussian distribution. Then, the detection  probability can be expressed as
\begin{align}\label{eq:P_D}
        P_{\rm D}&=1-\Phi\left(\frac{\tau_0-\mu_{u,1}}{\sigma_{u,1}}\right).
\end{align}
Note that the detection probability indicates the probability that the legitimate user is authenticated.

\subsection{Impersonation Procedure of Eve}

Eve intends to guess the secret key between the BS and the legitimate user. After Eve decodes the date of the legitimate user from BS correctly, Eve tries to detect the key from the maximum likelihood (ML) decoder. We assume the worst case scenario that Eve is able to decode all users' data from the BS. For the first step, Eve subtracts the contribution of users' data and tags, as
\begin{align}
    \textbf{R}_{\rm e}^{\rm H}&=\sqrt{\mathsf{P}_\mathsf{Tx}\mathsf{P}_\mathsf{t}}\textbf{H}_{\rm e}^{\rm H}\textbf{w}_u\textbf{t}_u^{\rm H}+\sum_{i=1}^{Z}\sqrt{\mathsf{P}_\mathsf{Tx}}\textbf{H}_{\rm e}^{\rm H}\textbf{v}_i\textbf{z}_i^{\rm H}+\textbf{N}_{\rm e}^{\rm H}.
\end{align}  
Eve estimates the tag applying LS equalization,
\begin{align}
    \hat{\textbf{t}}_{\rm e}^{\rm H}&=\textbf{t}_u^{\rm H}+\sum_{i=1}^{Z}\frac{\textbf{w}_u^{\rm H}\textbf{H}_{\rm e}\textbf{H}_{\rm e}^{\rm H}\textbf{v}_i}{\sqrt{\mathsf{P}_\mathsf{t}}\|\textbf{H}_{\rm e}^{\rm H}\textbf{w}_u\|^2}\textbf{z}_i^{\rm H}\nonumber\\
    &+\frac{1}{\sqrt{\mathsf{P}_\mathsf{Tx}\mathsf{P}_\mathsf{t}}\|\textbf{H}_{\rm e}^{\rm H}\textbf{w}_u\|^2}\textbf{w}_u^{\rm H}\textbf{H}_{\rm e}\textbf{N}_{\rm e}^{\rm H}~.
\end{align}
We assume that Eve has unlimited computing capability so that Eve generates all possible tags from the keys space. Eve obtains the expected key using a maximum likelihood (ML) decoder as follows:
\begin{align}
    \tau_{\rm e}&=\Re(\hat{\textbf{t}}_{\rm e}^{\rm H}\Tilde{\textbf{t}}_i),\\
   \eta_i&=\arg\max_{\eta\in\mathcal{K}}\tau_{\rm e}(\eta),
\end{align}
where $\mathcal{K}$ is the key space, and $\Tilde{\textbf{t}}_i$ is the i-th generated tag from the key space. 

Since Eve does not know any information of the correct key, Eve faces $|\mathcal{K}|$-ary hypothesis test. Once again, we approximate the distribution of $\tau_{\rm e}$ as Gaussian by the central limit theorem. Eve faces two hypotheses, whether the wrong key is chosen from the key space or the correct key is chosen from the key space. The means and variances on both hypotheses can be derived as
\begin{align}
    \mathbb{E}\{\tau_{\rm e}|H_0\}&=\mu_{\rm e,0}=0,\label{eq:H_0_m_2}\\
    \text{Var}\{\tau_{\rm e}|H_0\}&=\sigma_{\rm e,0}^2=\frac{L_{\rm t}}{2}\left(1+\sum_{i=1}^{Z}\frac{|\textbf{w}_u^{\rm H}\textbf{H}_{\rm e}\textbf{H}_{\rm e}^{\rm H}\textbf{v}_i|^2}{\mathsf{P}_\mathsf{t}\|\textbf{H}_{\rm e}^{\rm H}\textbf{w}_u\|^4}\right.\nonumber\\
    &\left.+\frac{1}{\mathsf{P}_\mathsf{Tx}\mathsf{P}_\mathsf{t}\|\textbf{H}_{\rm e}^{\rm H}\textbf{w}_u\|^2}\sigma_n^2\right),\label{eq:H_0_v_2}\\
    \mathbb{E}\{\tau_{\rm e}|H_1\}&=\mu_{\rm e, 1}=L_{\rm t},\\
    \text{Var}\{\tau_{\rm e}|H_1\}&=\sigma_{\rm e, 1}^2=\frac{L_{\rm t}}{2}\left(2+\sum_{i=1}^{Z}\frac{|\textbf{w}_u^{\rm H}\textbf{H}_{\rm e}\textbf{H}_{\rm e}^{\rm H}\textbf{v}_i|^2}{\mathsf{P}_\mathsf{t}\|\textbf{H}_{\rm e}^{\rm H}\textbf{w}_u\|^4}\right.\nonumber\\
    &\left.+\frac{1}{\mathsf{P}_\mathsf{Tx}\mathsf{P}_\mathsf{t}\|\textbf{H}_{\rm e}^{\rm H}\textbf{w}_u\|^2}\sigma_n^2\right).\label{eq:H_1_v_2}
\end{align}
Then, we can derive the probability that Eve decodes the correct key by ML decoder \cite{perazzone2019fingerprint},
\begin{align}
    P_{K}&=\int^{\infty}_{-\infty}\Phi\left(\frac{\tau-\mu_{\rm e,0}}{\sigma_{\rm e,0}}\right)^{|\mathcal{K}|-1}\varphi\left(\frac{\tau-\mu_{\rm e,1}}{\sigma_{\rm e,1}}\right){\rm d}\tau,\label{eq:P_K}
\end{align}
where $\varphi(\cdot)$ is the probability density function (PDF) of Gaussian distribution.

\section{Data Precoder and AN Precoder Design}

We design precoders for the data and AN individually. We consider well-known multi-user precoders for the data precoder and the null-space precoder for AN precoder. By this way, the BS can transmit data to the desired users while corrupting the data toward Eve using AN.  

We consider the regularized zero-forcing (RZF) linear precoder for the data part, which is given by
\begin{align}
    \text{RZF:}\quad\Tilde{\textbf{W}} &=(\textbf{H}\textbf{H}^{\rm H}+\beta\textbf{I})^{-1}\textbf{H}
    \label{eq:precoder_rzf_unnormalized}
\end{align}
where $\textbf{H} \,{=}\, \left[ \textbf{h}_1 \dots \textbf{h}_K \right]$ is the aggregate channel matrix, and $\beta$ is the regularization parameter. The power normalized precoder by the uniform power allocation scheme is given by:
\begin{align} \label{eq:precoder_user_normalized}
    \textbf{w}_k&=\sqrt{\frac{\phi}{K\,\mathsf{tr}(\Tilde{\textbf{w}}_k^{\rm H}\Tilde{\textbf{w}}_k)}}\Tilde{\textbf{w}}_k,
\end{align}
where $\phi$ is the power splitting factor between [0, 1], which indicates power division between data precoder and AN precoder. Note that $\phi$ can adjust the ratio of the transmitted power between data and AN.

The null-space precoding of AN is designed so that the power of AN does not reach at the desired users, as: 
\begin{align} \label{eq:precoder_eve_unnormalized}
    \Tilde{\textbf{V}}&=\text{null}(\textbf{H}^{\rm H}),
\end{align}
and the power normalized precoder vectors are obtained as:
\begin{align} \label{eq:precoder_eve_normalized}
    \textbf{v}_i &= \sqrt{\frac{1-\phi}{Z\,\mathsf{tr}(\Tilde{\textbf{v}}^{\rm H}_i\Tilde{\textbf{v}}_i)}}\Tilde{\textbf{v}}_i,
\end{align}
where the normalization scheme is the same as that of the data precoder.

\section{Power Allocation for Authentication}\label{sec:proposal}
In this section, we propose a power allocation strategy by jointly designing the tag power ($\mathsf{P}_\mathsf{t}$) and the power splitting factor ($\phi$) based on the  analytical interpretation. In previous related works \cite{zhu2014secure}, \cite{nguyen2018secure}, the power splitting factor is optimized to maximize the secrecy rate, which is one of the design criteria to evaluate the secure transmission. However, our work is not focused on how much data rate Eve can achieve, but whether Eve can illegally succeed in the authentication procedure by detecting the correct key. In addition, in previous literature, the transmission tag power is designed only by the criteria that the tag power should be low, in order not to reduce the data rate to the legitimate user, and the tag transmit power is not optimized separately or jointly. In our proposed strategy, we consider not only  the legitimate user's authentication probability but also the correct key detecting probability to design the power allocation strategy. Before introducing the proposed strategy, we remind that the transmitted power is distributed to the three factors: tag ($\textbf{t}_u$), data symbols ($\textbf{s}_k$), and AN ($\textbf{z}_i$) as shown in the system model in Fig.~\ref{fig:illu2}, while the power of two precoder, data precoder and AN precoder, is splitted by $\phi$.

\subsection{Analysis of Authentication Probability}\label{sec:proposal_1}
The authentication probability of the legitimate user is given in \eqref{eq:P_D}, and the probability is decided by the mean and the variance of two hypotheses in \eqref{eq:H_0_m_1}-\eqref{eq:H_1_v_1}. the mean in both hypotheses only depends on the length of sequence $L_{t}$, and  $L_{t}$ is constant in the power allocation strategy. The second term and third term in parentheses in \eqref{eq:H_0_v_1} can be regarded as zero, since multi-user precoder eliminates the effect of other users' channel and AN precoder is null to users' channel. Then, the variance in  \eqref{eq:H_0_v_1} can be approximated as
\begin{align}
     \text{Var}\{\tau_b|H_0\}&\approx\frac{L_{\rm t}}{2}\left(1+\frac{1}{\mathsf{P}_\mathsf{Tx}\mathsf{P}_\mathsf{t}|\textbf{h}_u^{\rm H}\textbf{w}_u|^2}\sigma_n^2\right),\label{eq:H_0_v_1_aprox}
\end{align}
where the remaining second term in parentheses in~\eqref{eq:H_0_v_1_aprox} comes from the additive noise. 

We observe that the variance in \eqref{eq:H_0_v_1_aprox} would be always fixed if we allocate the tag power  $\mathsf{P}_\mathsf{t}$ and power splitting factor $\phi$ by making the power allocation factor $\psi$ is the same. Note that the exact transmitted power of tag is $\mathsf{P}_\mathsf{t}\mathsf{P}_\mathsf{Tx}\phi$. The power allocation factor $\psi$ that determines the authentication probability can be written as 
\begin{align}\label{eq:pw_all_factor}
     \psi = \mathsf{P}_\mathsf{t}\phi,
\end{align}
where $\psi$ is derived from the denominator of the second term in~\eqref{eq:H_0_v_1_aprox} by calculating the contribution to the power allocation. Note that $\phi$ comes from the fact that the allocated power to $\textbf{w}_u$ is $\frac{\phi}{K}$ from \eqref{eq:precoder_user_normalized}. Moreover, the variance in \eqref{eq:H_1_v_1} depends on the power allocation factor $\psi$ by the same derivation as well. From the observation, we can conclude with the following \textit{Proposition~1}.

\textit{Proposition 1:} The authentication probability ($P_{\rm D}$) depends on $\psi$. With fixed $\psi$, we can design different combinations of $\mathsf{P}_\mathsf{t}$ and $\phi$, and it influences the performance of  Eve's key detection probability ($P_{\rm K}$) and the sum rate of the legitimate users. In other words, we can design $P_{\rm K}$ with the fixed $P_{\rm D}$, which affects the sum rate of the legitimate users and the Eve's correct key detecting probability.

\subsection{Analysis of Correct Key Detection Probability}\label{sec:proposal_2}
The correct key detection probability by ML decoder in \eqref{eq:P_K} is determined by the means and variances in \eqref{eq:H_0_m_2}-\eqref{eq:H_1_v_2}. By the similar way above, the means are not changed by the power allocations. For the variance in \eqref{eq:H_0_v_2}, we observe the behavior of variance when the SNR goes to infinity:
\begin{align}
    \lim_{\sigma\to0}\text{Var}\{\tau_{\rm e}|H_0\}&=\frac{L_{\rm t}}{2}\left(1+\sum_{i=1}^{Z}\frac{|\textbf{w}_u^{\rm H}\textbf{H}_{\rm e}\textbf{H}_{\rm e}^{\rm H}\textbf{v}_i|^2}{\mathsf{P}_\mathsf{t}\|\textbf{H}_{\rm e}^{\rm H}\textbf{w}_u\|^4}\right),\label{eq:H_0_v_2_approx}
\end{align}
where the second term in parentheses comes from the AN. We observe that the detection probability cannot converge to $1$ as the SNR increases due to the AN. We also observe that the variance in \eqref{eq:H_0_v_2_approx} would be always fixed with the power allocation factor $\omega$ as follows:
\begin{align}
    \omega=\frac{(1-\phi)\phi}{\mathsf{P}_\mathsf{t}\phi^2}
    =\frac{(1-\phi)}{\mathsf{P}_\mathsf{t}\phi},\label{eq:omega}
\end{align}
where $\omega$ is derived from the second term in parentheses in \eqref{eq:H_0_v_2_approx} by calculating the contribution to the power allocation. Note that the allocated power to $\textbf{v}_i$ is $\frac{1-\phi}{Z}$ from \eqref{eq:precoder_eve_normalized}. The variance in \eqref{eq:H_1_v_2} also follows the same behavior. From these observations, we can conclude with the following \textit{Proposition 2}.

\textit{Proposition 2:} The correct key detection probability ($P_{\mathsf{K}}$) at the high SNR regime is decided by $\omega$. The convergence value of $P_{\mathsf{K}}$ is determined by the $\omega$, which is given by the allocate the power $\mathsf{P}_\mathsf{t}$, $\phi$. Furthermore, since we can find the same $\omega$ with the different combinations of $\mathsf{P}_\mathsf{t}$, $\phi$, the authentication probability $P_{\mathsf{D}}$ and the sum rate are varied with fixed $P_{\mathsf{K}}$.

\subsection{Power Allocation Strategy by Integrating $\psi$, $\omega$}\label{sec:integrating}
From \textit{Proposition 1} and \textit{Proposition 2}, we conclude that the authentication probability ($P_{\mathsf{D}}$) and the correct key detection probability ($P_{\mathsf{K}}$) are affected by the power allocation factors $\psi$ and $\omega$, and the relation between these two factors is:
\begin{align}\label{eq:two_factors}
    \omega&=\frac{(1-\phi)}{\psi},\quad \omega<\frac{1}{\psi}
\end{align}
where the inequality comes from the constraint $0<\phi<1$. From \eqref{eq:pw_all_factor}, \eqref{eq:omega}, \eqref{eq:two_factors}, if we initially choose the  $\psi$, $\omega$ values depending on the promising  authentication probability ($P_{\mathsf{D}}$) and the correct key detection probability ($P_{\mathsf{K}}$), we can obtain corresponding the tag power ($\mathsf{P}_\mathsf{t}$) and the power splitting factor ($\phi$). In this manner, we can optimize power allocation depending on the legitimate user's channel quality ($P_{\mathsf{D}}$) and the desired secrecy ($P_{\mathsf{K}}$).

\section{Numerical Results}
\label{sec:results}
\begin{table}[!t]
\renewcommand{\arraystretch}{1.1}
\caption{Simulation settings}
\label{table:settings}
\centering
\begin{tabular}{lc}
\hline
Parameter & Value \\
\hline\hline
Transmit power ($\mathsf{P}_\mathsf{Tx}$) & [-10, 50] dBm\\
Antenna element spacing ($d_{\rm s}$) & $\frac{\lambda}{2}$ \\
Number of antennas at UAV-BS ($N$) & $16$ \\
Number of antennas at Eve ($M$) & 6 \\
Number of users ($K$) & 6 \\
Number of multipaths ($L_\mathsf{p}$) & 10 \\
Angular spread ($\Delta$) & $10^{\circ}$ \\
AoD/AoA distribution & Laplace \\
UAVs horizontal distance distribution ($d_{\mathsf{H}}$) & $\mathcal{U}[10,100]$ m \\
UAVs vertical distance ($d_{\mathsf{V}}$) & 100 m \\
Eve distance distribution ($d_{\rm{e}}$) & $\mathcal{U}[50,100]$ m \\
Carrier frequency ($f_{\rm c}$) & 28 GHz \\
Bandwidth ($B$) & 100 MHz \\
Noise figure & 9 dB \\
Thermal noise & {-}174 dBm/Hz\\ 
Power splitting factor ($\phi$) & $[0, 1]$ \\
Tag power factor ($\mathsf{P}_\mathsf{t}$) & $[0, 1]$ \\
length of tag sequence ($L_{\rm t}$) & $2048$ \\
 false alarm probability ($p_{\rm fa}$) & 0.001\\\hline
\hline
\end{tabular}
\vspace{-0.15in}
\end{table}

\begin{figure}[!t]
	\centering
	\subfloat[Authentication probability.]{
	\includegraphics[width=0.5\textwidth]{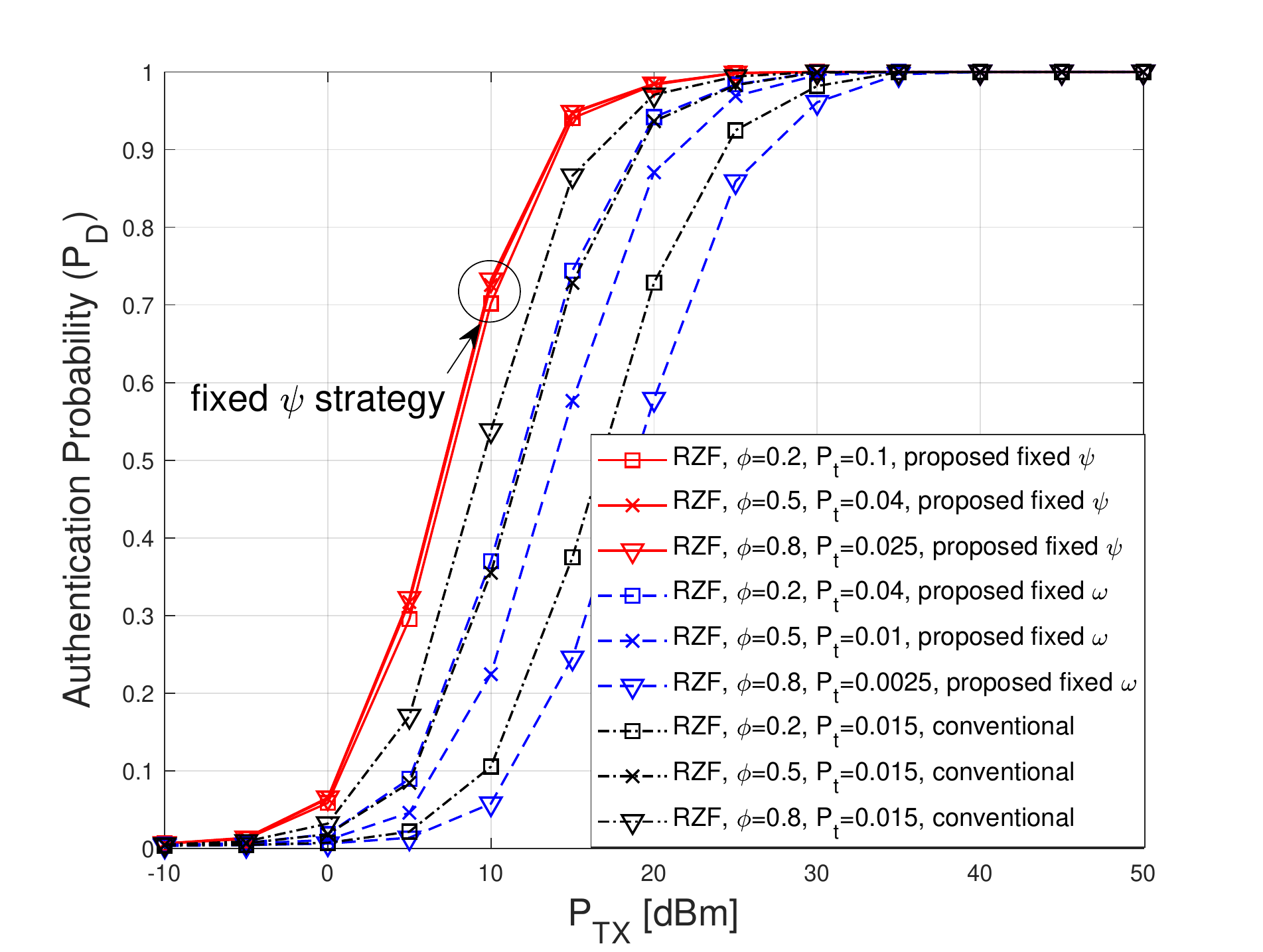}\label{fig:1_1}}
	\vspace{-0.0in}
	\subfloat[Correct key detection probability.]{
	\includegraphics[width=0.5\textwidth]{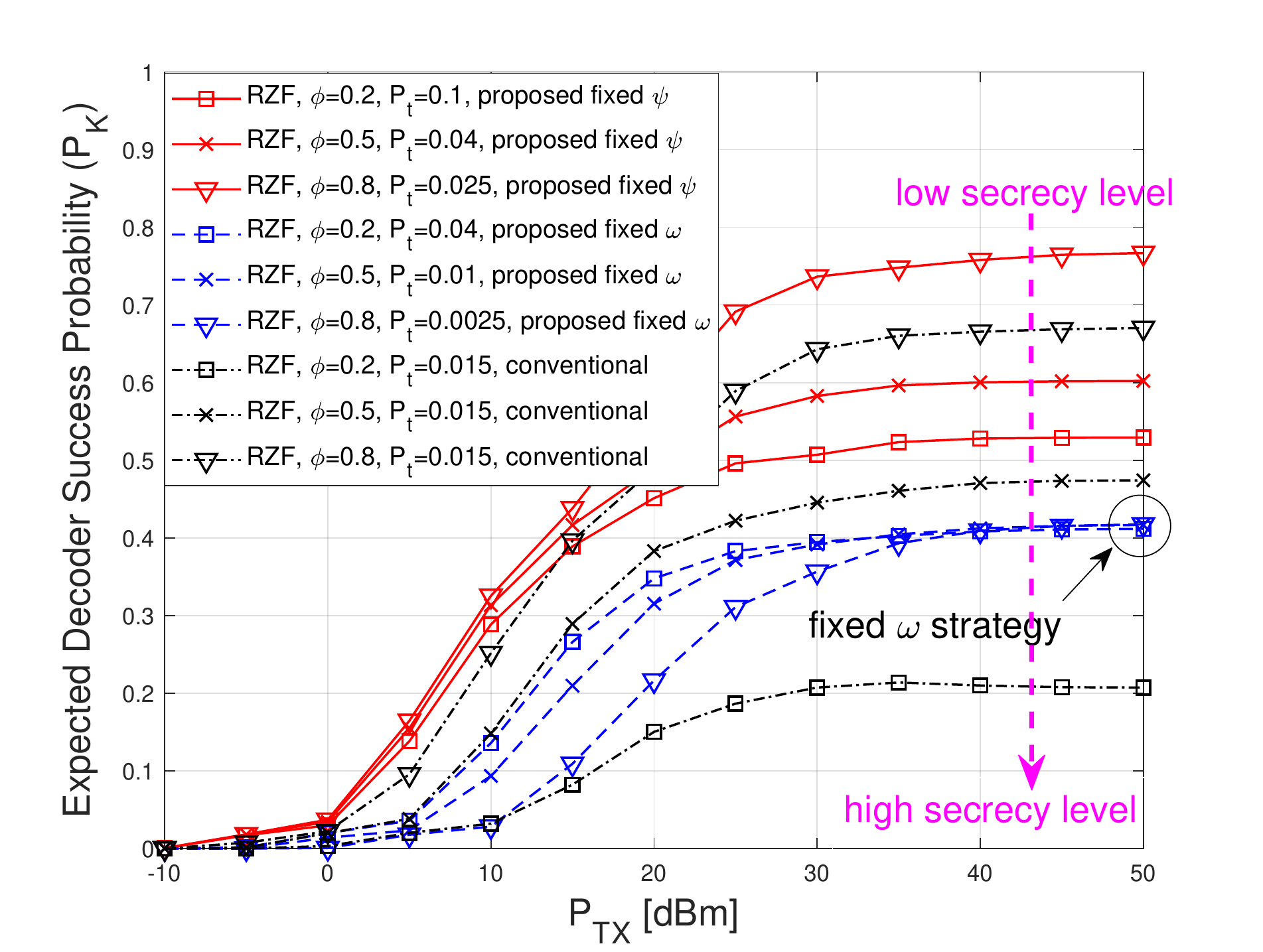}\label{fig:1_2}}
	\vspace{-0.0in}
	\subfloat[Sum rate.]{
	\includegraphics[width=0.5\textwidth]{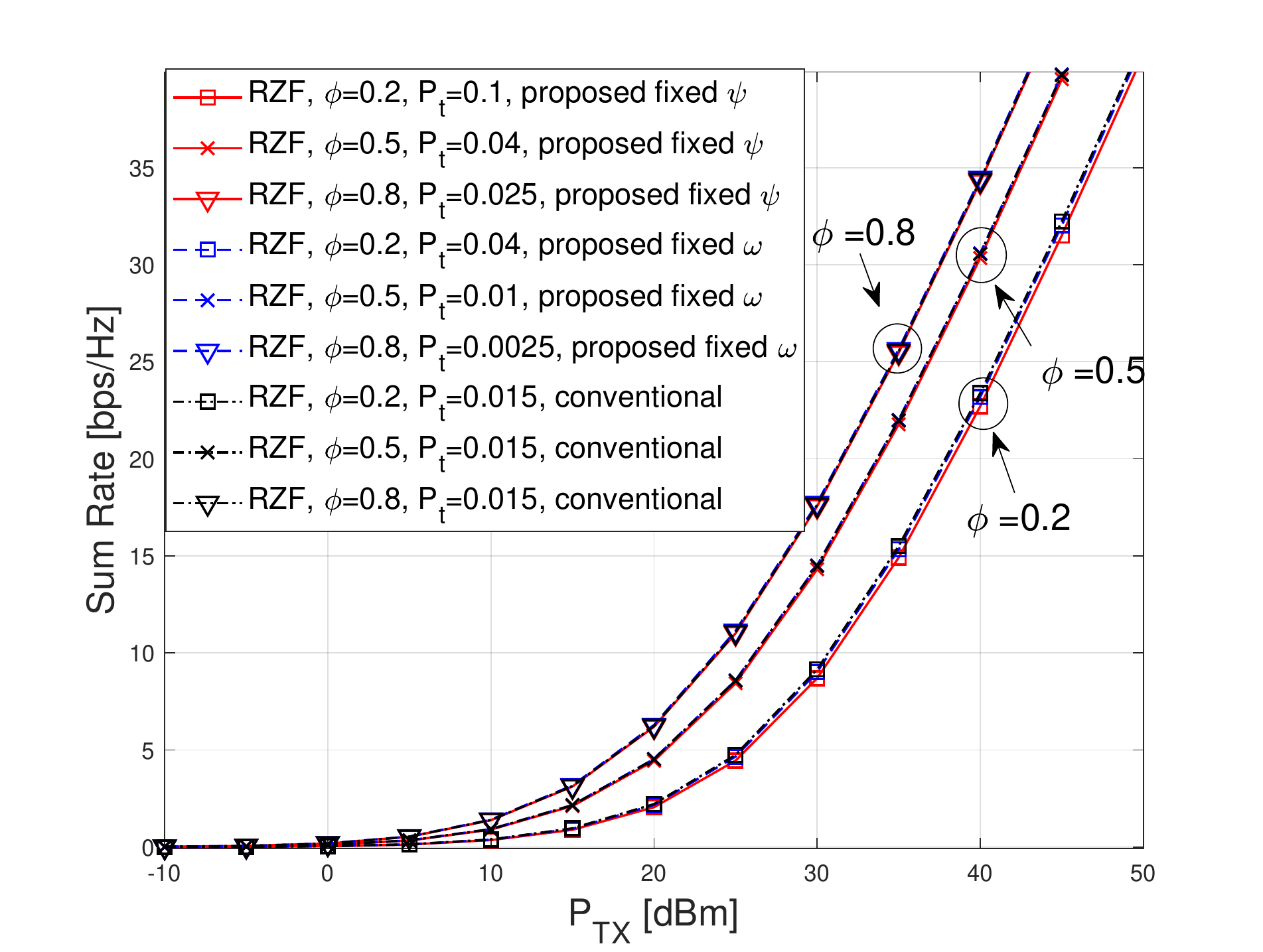}\label{fig:1_3}}
	\vspace{-0.0in}
	\caption{Performance evaluation of different power allocation strategies ($\psi = 0.02$, $\omega = 100$ for the proposed fixed strategies). } 
	\label{fig:1}
	\vspace{-0.35in}
\end{figure}

In this section, we present numerical results to evaluate the performance of the proposed power allocation strategies compared with the conventional approach. The simulation parameters are listed in Table~\ref{table:settings}.
Fig.~\ref{fig:1} compares the performance of three different  power allocation strategies by the authentication probability, the expected decoder success probability, and the sum rate. The red lines, the blue lines, and the black lines indicate the proposed fixed $\psi$ strategy, the proposed fixed $\omega$, and the conventional strategy, respectively. 

In the conventional strategy, the tag power is fixed as $0.015$ and only $\phi$ varies, which is adopted in  \cite{perazzone2019fingerprint}, \cite{perazzone2019physical}. First of all, we can observe that the authentication probability ($P_{\mathsf{D}}$) is the same by the proposed fixed $\psi$ strategy in Fig.~\ref{fig:1_1} and the expected decoder success probability ($P_{\mathsf{K}}$) converges to the same probability by the proposed fixed $\omega$ strategy in Fig.~\ref{fig:1_2}, which we discuss in Sections~\ref{sec:proposal_1}, \ref{sec:proposal_2}. Second, it is observed that the sum rate of legitimate users increases as $\phi$ increases, regardless of $\mathsf{P}_\mathsf{t}$ in Fig.~\ref{fig:1_3}. This is due to the fact that, increasing $\phi$ means that the allocated power to data increases and $\mathsf{P}_\mathsf{t}$ is sufficiently small not to degrade the sum rate. 

Third, when we increase $\phi$ in order to improve the sum rate in the conventional strategy, both authentication probability and the expected decoder success probability increase. However, when we increase $\phi$ in order to improve the sum rate in the proposed fixed $\psi$ and fixed $\omega$ strategies, authentication success level and secrecy level is stable by the respective strategies, which means that we can hold either authentication success or secrecy level even if we improve the sum rate by increasing $\phi$. Note that since we do not know the channel quality of Eve in general, we can regard the secrecy level as the converged probability of $P_{\mathsf{K}}$, which is the worst decoding probability. Besides, we can also interpret that we can improve the sum rate with the fixed secrecy level by the cost of the authentication probability of the legitimate user.

\begin{figure}[t]
	\centering
	\includegraphics[width=0.5\textwidth]{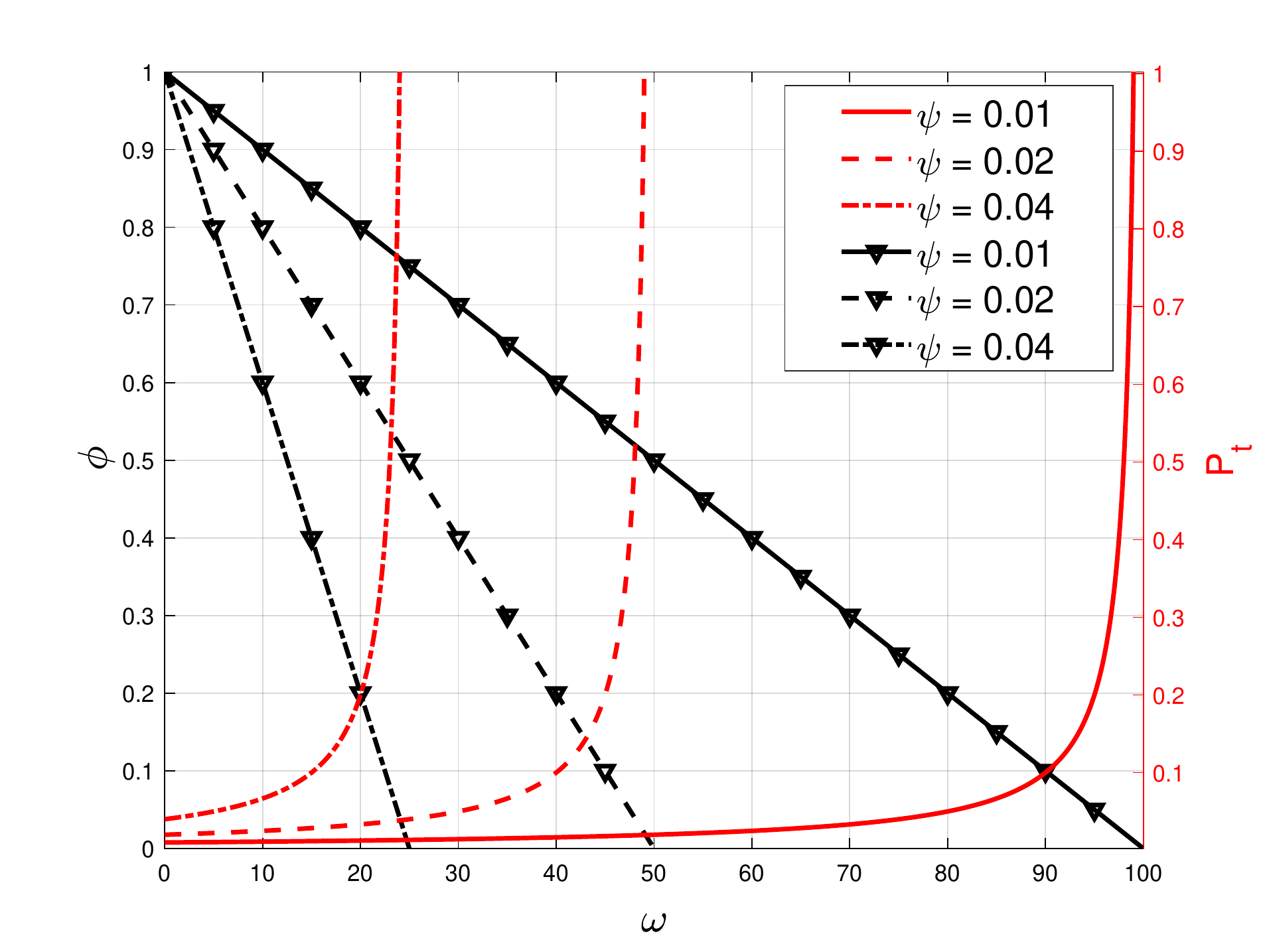}
	\caption{The relation between power allocation factors $\psi$, $\omega$ and $\phi$, $\mathsf{P}_\mathsf{t}$.} 
	\label{fig:2}
	\vspace{-0.1in}
\end{figure}

In Fig.~\ref{fig:2}, we show the relationship among the related parameters in the proposed strategy: $\psi$, $\omega$, $\phi$, and $\mathsf{P}_\mathsf{t}$. The relation can be obtained from \eqref{eq:pw_all_factor}, \eqref{eq:omega}, \eqref{eq:two_factors}. As discussed in Section~\ref{sec:integrating}, if we decide authentication success and secrecy level by selecting $\psi$ and $\omega$,  corresponding $\phi$, $\mathsf{P}_\mathsf{t}$ are determined as well. It is observed that as $\omega$ increases, $\phi$ increases and $\mathsf{P}_\mathsf{t}$ decreases. It means that sum rate decreases as secrecy level increases with the fixed authentication probability. We also observe that we can increase $\phi$ by decreasing $\psi$ with the fixed $\omega$. This observation interprets that sum rate increases by reducing authentication success level with the same secrecy level.

\section{Conclusion}\label{sec:conclusion}
In this paper, we study the physical layer
authentication via fingerprint embedding with artificial noise. We consider multi-user scenario with mmWave  communications, where we adopt  regularized zero-forcing precoding for data and  null-spacing precoding for artificial noise. We propose the power allocation strategies, to determine the power splitting factor and the tag power factor, which control the allocated power among data, artificial noise, and tag. We show that we can manage the authentication success level on the legitimate users' side as well as the secrecy level on the Eve's side by the proposed power allocation strategies. We also compare the proposed strategies with the conventional approach, which fixes tag power.
\vspace{-0.1in}

\bibliographystyle{IEEEtran} 
\bibliography{IEEEabrv,bibfile}

\end{document}